\documentclass{elsart3p}

\usepackage{natbib}

\usepackage{graphics}
\usepackage{epsfig}

\usepackage{amssymb}

\newcommand{\beq}{\begin{equation}}
\newcommand{\eeq}{\end{equation}}
\def\la{\hbox{\raise.35ex\rlap{$<$}\lower.6ex\hbox{$\sim$}\ }}
\def\ga{\hbox{\raise.35ex\rlap{$>$}\lower.6ex\hbox{$\sim$}\ }}

\def\beq{\begin{equation}}
\def\eeq{\end{equation}}
\def\beqa{\begin{eqnarray}}
\def\eeqa{\end{eqnarray}}

\def\order#1{{\cal O}\left({#1}\right)}
\newcommand{\sfrac}[2]{ \mbox{$\frac{#1}{#2}$}}

\begin{document}

\begin{frontmatter}



\title{Hydrodynamical activity in thin accretion disks}


\author[lab1,lab2]{Oded Regev}

\ead{regev@physics.technion.ac.il}


\address[lab1]{Department of Physics, Technion - Israel Institute of Technology}
\address[lab2]{Department of Astronomy, Columbia University}

\begin{abstract}
An asymptotic treatment of thin accretion disks, introduced by Klu\'zniak \& Kita (2000)
for a steady-state disk flow, is extended to a time-dependent problem. Transient growth of
axisymmetric disturbances is analytically shown to occur on the global disk scale. The implications
of this result on the theory of hydrodynamical thin accretion disks, as well
as future prospects, are discussed.
\end{abstract}


\end{frontmatter}

\section{Introduction}
\label{1sec}
Thin accretion disks (AD) form whenever a sufficiently cool gas, endowed with a
significant amount of angular momentum, is gravitationally attracted
towards a relatively compact object. This situation is quite common in astrophysics
and therefore the observational and theoretical study of accretion
disks has been quite intensive. The necessary condition for accretion
to take place is that angular momentum be extracted from the fluid
swirling around the central object. To achieve accretion rates that
are consistent with observations, the physical mechanism for such angular momentum
transport must be more efficient (by many orders of magnitude)
than just the one resulting from torques caused by microscopic viscosity.

Already at the outset, when AD were theoretically proposed
(Prendergast \& Burbidge 1968, Pringle \& Rees 1972), the enhanced
effective viscosity was postulated to result from turbulence and
was parametrized, using mixing length theory or a similar scheme,
as detailed theoretical understanding of turbulence and the
transition to it was lacking then (a situation that is still with
us nowadays). The parametrization of the effective viscosity in
disks in terms of a single parameter - $\alpha$, introduced by
Shakura \& Sunyaev (1973), proved itself to be the most fruitful,
giving rise to successful interpretations of many observational
results (see Lin \& Papaloizou 1996, Frank, King \& Raine 2002,
for reviews). In some cases (e.g. dwarf nova models), however,
more complex (and cumbersome) prescriptions for the viscosity had
to be employed. However, until the early 1990's the question what
is the physical origin of turbulence (or, more precisely, the
anomalous angular momentum transport) in AD has essentially
remained unanswered. No definite linear instability has been
identified for thin disks in which the swirling flow consist of
Keplerian shear. The magneto-rotational instability (MRI),
originally found (Velikhov 1959, Chandrasekhar 1960) for magnetic
Taylor-Couette, (i.e., cylindrical) flows, has been shown by
Balbus \& Hawley (1991) to also operate in cylindrical AD, when
the fluid is electrically conducting and for not too large initial
magnetic fields.

It has thus become the operating paradigm in
the astrophysical community that purely hydrodynamical turbulence
in Keplerian disks is altogether ruled out and since the problem
of angular momentum transport in these objects relies on MRI driven
magneto-hydrodynamical (MHD) turbulence, extensive numerical calculations
are necessarily the main tool of research on this problem
(see Balbus \& Hawley 1998 and Balbus 2003, for reviews and references).
Nonetheless, efforts to find a purely hydrodynamical transition to
turbulent activity in thin AD have continued until the present
time, more than a decade after Balbus , Hawley \& Stone (1996)
appeared to settle the matter. These have largely been motivated by the
fact that a purely hydrodynamical AD flow has a `microscopic' Reynolds
number ($\rm Re$) of the order of $10^{14}$
or so, a quite amazing setting for a laminar flow.

Among the ideas that have been put forward in this context
some are based on the application to thin disk flows of the
viewpoint, familiar to the fluid-dynamical community, that {\em
transient dynamics induced by perturbations}, i.e. strong transient growth
(TG) in linearly stable shear flows may play an important r\^ole in nonlinearly
shaping the final dynamical state. TG is possible because the relevant linear
operator governing the behavior of infinitesimal perturbations
in these flows is non-normal and thus the usual modal approach
essentially fails (see, e.g., Grossman 2000, Schmid \& Hennigson 2001, and
Criminale, Jackson \& Joslin 2003)

TG of disturbances has been discussed
in the astrophysical literature within two quite different (but related) contexts:
\begin{description}
\item[{$\bullet$}] Local disturbances experiencing large enough
TG that can possibly trigger a subcritical nonlinear transition
into turbulence in a linearly stable shear
flow, i.e. via the so-called bypass transition (e.g., Chagelishvili {\em et} al. 2003,
Tevzadze {\em et} al. 2003, Yecko 2004, Afshordi {\em et} al. 2005, Mukhopadhyay {\em et} al. 2005)
\item[{$\bullet$}]
Perturbations experiencing significant TG that can be excited either by some external
agent, perhaps as secondary flows on a pre-existing 3D turbulence itself, however weak,
may give rise to intense global dynamical activity (e.g. Ioannaou \& Kakouris 2001,
and see below).
\end{description}
Other scenarios for inducing hydrodynamical activity in AD
(some of them directly or indirectly related to TG as well)
have also been proposed. Among them are those invoking baroclinic instabilities
(e.g., Klahr \& Bodenheimer 2003), strato-rotational instabilities (e.g., Dubrulle
{\em et} al. 2005, Umurhan 2006), formation and long sustenance of vortices and/or waves
(e.g., Godon \& Livio 1999, Bracco {\em et} al. 1999, Li {\em et} al. 2001,
Umurhan \& Regev 2004, , Barranco \& Marcus 2005, Petersen,
Stewart \& Julien 2007, Bodo {\em et} al. 2007, Lithwick 2007). Even though no definite
and undisputed hydrodynamical mechanism, that can give rise to sufficient angular momentum
transport in AD, has been identified so far, significant efforts along these
lines are continuously being made by several groups of researchers.

It appears that continued study of purely hydrodynamical
processes in disks remains viable and worthwhile, in particular in view of the
difficulties with MRI driven angular momentum transport in AD, which have recently been
pointed out. Questions of insufficent numerical resolution in MHD disk simulations have
been convincingly raised (Pesah, Chen \& Psaltis 2007, Fromang \& Papaloizou 2007).
The decrease of transport with decreasing magnetic Prandtl number (${\rm Pm}$) for various
setups and boundary conditions and, in particular, the vanishing of this transport for
${\rm Pm}\ll 1$ (which is the case in many instances of astrophysical AD) has been
demonstrated (Umurhan, Menou \& Regev 2007, Lesur \& Longaretti 2007, Umurhan, Regev \& Menou 2007,
Fromang {\em et} al. 2007). Finally, serious doubts as to the viability of the local shearing box
approximation (Goldreich \& Lynden-Bell 1965) to the numerical study of accretion disk MHD turbulence
(Coppi \& Keyes 2003, King, Pringle \& Livio 2007, Shu {\em et} al. 2007,
Regev \& Umurhan 2007) have been raised.

The purpose of this contribution is to draw attention to the possibility
that a thin AD, in which a sub-critical transition to
hydrodynamical turbulence occurs (for a sufficiently high ${\rm Re}$)
large transient amplification of global disturbances may
induce recurrent or even persistent secondary flow activity.
Lesur \& Longaretti (2005) have explicitly
demonstrated, using high resolution 3D numerical simulations,
that such sub-critical transition to turbulence
does appear in a model flow having Keplerian shear characteristics,
but at a very high
${\rm Re}$ (see also Rincon {\em et} al. 2006).
However, it appears that the efficiency of turbulent
transport in these flows is insufficient for astrophysical
purposes (i.e. AD). A similar conclusion can perhaps be drawn
from the recent experimental study of Ji {\em et} al. (2006). The very
small value of the effective $\alpha$ (measuring the angular momentum
transport) in these flows is, as we shall show, the key necessary ingredient for
the excitation of vigorous global (transient) secondary flows, atop the weak turbulent
state.
It is quite conceivable that the large (by orders of magnitude in the
disturbance energy) TG may (nonlinearly) give rise to persistent
dynamical activity, or at least be recurrently re-excited by external
perturbations (see Ionnaou \& Kakouris 2001, who advocated the latter possibility).
It however remains to be shown, of course, that in the ultimate state angular momentum
transport is appropriate for AD, i.e. the effective $\alpha$
acquires a high enough value.

The primary tool utilized here in order to facilitate
an analytical treatment is the asymptotic expansion, where the
dependent variables and governing equations are expanded in powers
of a small quantity (here the measure of the disk's `thickness',
$\epsilon$ - see below).
Exposing the resulting mathematical system (an initial value
problem) to a set of initial conditions, in which the extreme geometry
of the disk structure is taken into account (i.e. $\epsilon \ll 1$),
we  aimed at obtaining an analytically treatable problem. To achieve
this goal we also assumed a polytropic relation between the
pressure and density.
The advantage of such an approach is obvious - the
treatment can be essentially analytical and the responsible physical effects
leading to any interesting dynamics may be transparently traced.
The study reported on here is limited to axi-symmetric disturbances
and the question of its ultimate development still remains open.
We are now in the process of generalizing it to fully 3-D disturbances
and it is clear that the present analytic work should be
ultimately complemented with detailed and uncompromising 3-D numerical
calculations with a proper treatment of energy generation and transfer.

\section{Asymptotic formulation of the dynamics of a thin polytropic accretion disk}
\label{2sec}

We wish to investigate the dynamical (i.e. time dependent) behavior of thin,
axisymmetric AD, allowing for vertical structure.
{\em Viscous} flow is invoked, with a standard viscosity prescription, that is
we actually assume that an effective viscosity producing
mechanism is already operative. The mean flow is thus assumed to consist
of Keplerian thin disk flow, which has already undergone a sub-critical
transition, like the one found by Lesur \& Longaretti (2005), for example.
As mentioned above, these authors found that the value of the effective $\alpha$
scales with the value of the sub-critical transition Reynolds number, ${\rm Re}_g$,
as $\alpha \sim 1/{\rm Re}_g$ and since they obtained that the
value of ${\rm Re}_g$ for Keplerian rotation flows is extremely high, they
concluded that the $\alpha$ values that can be expected are much smaller than
those necessary for angular momentum transport in astrophysical AD.
He we shall examine the dynamical behavior for different values of $\alpha$ and,
as we shall see, the TG we find will be anti-correlated with the value of
$\alpha$ of the mean steady flow.

The key observation is that when on writes the equation of fluid dynamics (e.g.,
Tassoul 1978) and makes them non-dimensional by scaling the variables with their
typical values, a non-dimensional parameter $\epsilon$
appears, multiplying (by its
value in various powers) different terms. These (axisymmetric) equations in
cylindrical coordinates ($r, z, \phi$) read:
\beq
\partial_t\rho + \frac{\epsilon}{r}\partial_r(r\rho u)
+ \partial_z (\rho v)= 0,
\label{continuity-eqn}
\eeq
\beqa
&&
\epsilon {\partial_t u} + \epsilon^2 u {\partial_r u}
+ \epsilon v{\partial_z u} - \Omega^2r =
- \frac{1}{r^2}\left(1 + \epsilon^2\frac{z^2}{r^2}\right)^{-{3 \over 2}}
\nonumber\\
&&
+ {\epsilon \over \rho} \partial_z\left(\eta {\partial_z u}\right)
+ {\epsilon^2 \over \rho} \left[\rho \partial_r \varpi + \partial_z
\left(\eta \partial_r v \right) - \frac{2}{3} \partial_r
\left(\eta \partial_z v\right) \right]\nonumber\\
&&
 + {\epsilon^3 \over \rho}\left\{-\frac{2\eta u}{r^2} + \frac{2}{r}{\partial_r}
\left(\eta r {\partial_r u}\right)  - \frac{2}{3}{\partial_r}
\left[{\eta \over r}\partial_r(ru)\right]   \right\},
\label{u-eqn}
\eeqa
\beqa
&&
\partial_t v + \epsilon \partial_r v +v \partial_z v =
- \frac{z}{r^3}\left(1 + \epsilon^2\frac{z^2}{r^2}\right)^{-{3 \over 2}}-
\partial_z \varpi
\nonumber\\
&&
+ \frac{4}{3\rho}\partial_z\left(\eta \partial_z v \right)
+ \frac{\epsilon}{\rho r}\left\{
\partial_r \left(\eta r \partial_z u\right)
- \frac{2}{3}\partial_z\left[{\eta}\partial_r (ru)\right]
\right\}
\nonumber\\
&&
~~~~ +\epsilon^2\frac{1}{\rho r}
\partial_r\left(\eta r \partial_r v \right),
\label{v-eqn}
\eeqa
\beqa
\partial_t\Omega + \epsilon\frac{u}{r^2}\partial_r (r^2 \Omega)
+v\partial_z\Omega &&= \nonumber\\
+{1 \over \rho}\partial_z\left(\eta \partial_z \Omega \right)
&& +\frac{\epsilon^2}{\rho r^3}\partial_r
\left(\eta r^3 \partial_r \Omega\right).
\label{Omega-eqn}
\eeqa
In these equations $u, v$ and $\Omega$ are the horizontal, vertical
and angular velocity respectively, $\varpi \equiv n c_s^2 = (n+1)\rho^{1/n}$
($c_s$ is the sound speed and the non-dimensional polytropic
relation $P=\rho^{1+1/n}$, with the general polytropic index $n$,
has been used). Also, $\eta$ is the dynamic viscosity, which can be expressed
using the standard Shakura-Sunyaev viscosity prescription
\beq
\eta = {2 \over 3}{\alpha P \over \Omega_{\rm K}} =
{2 \over 3}\alpha~\rho^{1+1/n}~r^{-3/2},
\label{SSvis}
\eeq
with the non-dimensional Keplerian angular velocity substituted as $\Omega_{\rm K}=r^{-3/2}$.

The non-dimensional parameter $\epsilon \equiv \tilde c_s/(\tilde \Omega \tilde r) = \tilde h/\tilde r$
(the quantities denoted by tilde are the corresponding dimensional quantities, evaluated
at a typical position in the AD, with $\tilde h$ being the disk height)
is very small for thin disks (that is, those assumed to be able to cool
efficiently, so that the rotational velocity is highly supersonic).

An asymptotic approach of the kind used here was introduced for the
first time to the study of thin viscous AD by Regev (1983),
in the context of AD boundary layers and later was developed and
used in a remarkable analytical work of Klu\'zniak \& Kita (2000),
(hereafter KK) who solved for the {\em steady} structure of a
polytropic viscous
axisymmetric disk. This study revealed the presence of a steady
meridional flow pattern with backflow for values of the $\alpha$
less than some critical value. This result and feature was
confirmed by Regev \& Gitelman (2002), who abandoned the
polytropic assumption and included an energy equation (employing
the diffusion approximation in the treatment of the vertical
radiative transport), showing that
the polytropic assumption makes only very little substantive difference
from the {\em steady} meridional flow solution of KK.

The next step of the procedure consists of expanding all dependent variables
asymptotic series in $\epsilon$, for example
\beq
\Omega= \epsilon \Omega_0 + \epsilon^2 \Omega_2 + \epsilon^4 \Omega_4...
\eeq
Similar expansions are used for $\rho$ and $v$, while the one for $u$
consists of only odd powers of $\epsilon$. The feasibility of this choice
is justified in Umurhan {\em et} al. (2006) (hereafter UNRS). These expansions are then
substituted into the
full equations (\ref{continuity-eqn}-\ref{Omega-eqn}) and the
expression in various orders in $\epsilon$ are collected. The resulting equations are similar
to the ones found by KK, but there are some important
differences:
\begin{enumerate}
\item
Time-dependence is included so as to be able to study the dynamics and consequently
time-derivatives of the dependent variables appear (the time unit taken as the
typical rotation time, $\tilde t = 1/\tilde \Omega$). The problem is thus formulated
as an {\em initial value} problem.
\item
Time-dependence is excluded in the lowest order term in the expansions
and it is introduced only in the next significant order,
in the form of an {\em additive} term, denoted by
a prime, and can be viewed as a perturbation on the steady-state, but not necessarily
an infinitesimal one)
Thus the typical even and odd order expansions
are of the form
 \beqa
 && \rho(r,z,t)  =  \rho_0(r,z)+\epsilon^2\left[\rho_2(r,z)
 +\rho_2'(r,z,t)\right]\nonumber\\
 &&~~~~~ + \epsilon^4\left[  \rho_4(r,z) +\rho_4'(r,z,t)\right]+\cdots
 \label{expan-even}\\
 && u(r,z,t)  =  \epsilon \left[u_1(r,z) + u_1'(r,z,t)\right]+\nonumber\\
 &&~~~~~~~~ +  \epsilon^3 \left[ u_3(r,z) +u_3'(r,z,t)\right]+ \cdots
 \label{expan-odd}
 \eeqa
 \item
 As in KK the solution of $\order1$ equations yield a steady disk of the Shakura
 \& Sunyaev kind, but for a polytrope (see also UNRS for the case of an arbitrary
 polytropic index $n$). The $\order\epsilon^2$ equations can be conveniently split into
 a steady and time-dependent parts, owing to the additive construction of the
 time-dependent disturbances. The steady part yields the KK steady solution,
 with possible backflows (which occur for $\alpha<\alpha_c\sim 0.7$).
 \end{enumerate}

 We conclude this section by formulating the initial
 value problem resulting from the $\order\epsilon^2$ time-dependent equations.
\beqa
 \partial_t \rho'_2 &=& -\frac{1}{r} \partial_r ( r \rho_0 u'_1)
- \partial_z (\rho_0 v'_2),
\label{ivp1}\\
\partial_t u'_1 &=& 2 \Omega_0 \Omega'_2 r +
\frac{1}{\rho_0} \partial_z \left( \eta \partial_z u'_1 \right),
\label{ivp2}\\
\partial_t v'_{2} &=& -\partial_z \varpi'_2 +\frac{4}{3 \rho_0}
\partial_z \left(\eta \partial_z v'_2\right) +
\frac{1}{\rho_0 r}\partial_z \left(\eta r\partial_z u'_1 \right)
\nonumber\\
&&~~~~~~~~~~~ -  \frac{2}{3\rho_0}
\partial_z\left\{\eta
 \left[\frac{1}{r}\partial_r (ru'_1)\right]\right\},
\label{ivp3}\\
 \partial_t \Omega'_2 &=& -\frac{ u'_1}{r^2} \frac{d (r^2\Omega_0)}{d r}
 + \frac{1}{\rho_0}\partial_z\left(
 \eta \partial_z \Omega'_2\right),
\label{ivp4}
\eeqa
where the zero-indexed quantities are known from the steady $\order1$
(Shakura-Sunyaev disk) solution (e.g., $\Omega_0=r^{-3/2}$) and we base the
expression for $\eta$ on the $\order1$ solution only, which using
(\ref{SSvis}) is
$\eta = (2/3)\alpha~\rho_0^{1+1/n}~r^{-3/2}$.
In addition, from the polytropic law it follows that
$\varpi'_2 = \varpi_0 \rho'_2/\rho_0$.

The initial value problem
defined by (\ref{ivp1}-\ref{ivp4}) must be complemented by boundary
conditions. We have opted for retaining the integral steady mass inflow condition
as in KK. In addition, guided by the physical consideration that there should be
no work done on the disk surface, we have required that the external pressure
and viscous stresses vanish at this surface. For details see UNRS.

\section{Representative solutions}
\label{3sec}
Consider now solutions to the initial value problem, posed
in the previous section and for the sake of simplicity we set,
from here and on, the
polytropic index to the value $n=3/2$ (as in KK). All solutions
obviously depend on the initial conditions and
our purpose here is to discuss some relevant representative solutions,
that can exhibit TG. It is important to note that the above
four equations support two types of solutions, namely a pair comprising
of purely vertical disturbances and a second pair containing disturbances
in all the velocity components. This follows from the fact that equations
(\ref{ivp2}) and (\ref{ivp4}) dynamically decouple from the
other two,  i.e., the quantities
$u'_{1}$ and $\Omega'_{2}$ are not dynamically influenced by
$v'_{2}$ and $\rho'_{2}$, while the converse is not true. Thus,
in principle, if perturbations are chosen such that $u'_{1} = \Omega'_{2} =0$
initially, they will remain so at all times and the motion will
be of purely vertical acoustics (VA), as embodied in $v'_{2}$ and $\rho'_{2}$.
On the other hand, if one chooses non-vanishing
$u'_{1}~~ \&~~ \Omega'_{2}$ as initial conditions, the resulting dynamics
comprises of all variables. We shall call
these solutions "driven" general acoustics (GA).
This mathematical structure
is reminiscent of the behavior of the steady-state equations of KK
and we shall now elaborate on these two solution types.

\subsection{Meridional velocity disturbances}
It is quite straightforward to derive from the
differential set (\ref{ivp1}-\ref{ivp4})
two second order equations for just the function $u'_{1}(r,z,t)$ and $v'_{2}(r,z,t)$,
that is, the lowest order time-dependent meridional components of the velocity
perturbations. They are
\beqa
{\cal P} u'_{1} &=& 0,
\label{ueq}\\
{\cal L}v'_{2}&=& \left[\partial_t {\cal F} +  {\cal G} \right]u'_{1},
\label{veq}
\eeqa
where the differential operators are defined (as operating on an arbitrary function
$\varphi(r,z,t)$) in the following way

\beqa
{\cal P} &\equiv&
\left[\partial_t  - \frac{1}{\rho_0}\partial_z (\eta \partial_z)
\right]\left[\partial_t\varphi  - \frac{1}{\rho_0}\partial_z (\eta \partial_z\varphi)
\right]+\nonumber\\
&&~~~~~~~~~~~~~~~~~~~~~~~~~~~~~~~~~~~~~~~
 + \Omega_0^2 \varphi,
\nonumber\\
{\cal L} &\equiv&
\partial_t^2 \varphi  - {2 \over 3}\varpi_0 \partial^2_z \varphi
-{5 \over 3} \partial_z \left( \varpi_0 \partial_z \varphi \right)
- \left(\partial^2_z  \varpi_0\right) \varphi -\nonumber\\
&&~~~~~~~~~~~~~~~~~~~~~
-\frac{4}{3\rho_0} \partial_z \left[
\eta \partial_z \left( \partial_t \varphi  \right) \right],
\nonumber\\
{\cal F} &\equiv&
-\frac{2}{3\rho_0 }\partial_z [\eta \partial_r ( r \varphi)]
+
\frac{1}{r\rho_0}\partial_r(r\eta \partial_z \varphi),
\nonumber\\
{\cal G} &\equiv& \partial_z\left[
\frac{2 \varpi_0}{3 r\rho_0}\partial_r ( r \rho_0  \varphi)\right].
\label{operators}
\eeqa
The boundary conditions completing this system can be specified
using the `free boundary' prescription at the disk surface, as
mentioned at the end of the previous section (see UNRS for an explicit
formulation). After solutions to equations (\ref{ueq}-\ref{veq}) are
found, it is possible to return to the original set of equations
in order to find the two additional unknown functions
$\rho'_2(r,z,t)$ and $\Omega'_2(r,z,t)$.

We were able to find analytical solutions to the
system (\ref{ueq}-\ref{veq}). In describing them we shall first
discuss, in subsection \ref{homo}, the case $u_1'=0$ and
$\Omega_2'$=0, which result from the special initial conditions
in which these perturbations are not excited.
This will thus allow us to be limited to solving only the homogeneous part of
(\ref{veq}). The solutions of this equation
constitute the above mentioned first pair of modes - VA.
The solutions to the full set (\ref{ueq}-\ref{veq}) in
the general (i.e. inhomogeneous case, with $u_1' \neq 0$)
will be discussed in subsection \ref{general}. This
pair of modes was called before GA and we remark once again that
their dynamics can be viewed as being driven (by the inhomogeneous part).
Finally, in subsection \ref{TG} we shall describe in some detail
the TG behavior found in our solutions.
A full exposition of the calculations leading to the solutions quoted and
discussed in subsections \ref{homo}-\ref{TG} is presented in UNRS
and its Appendices. In the
following three subsections only the salient features will be given.

\subsection{Vertical acoustics (VA)}
\label{homo}
The general solution to the inhomogeneous equation (\ref{veq}) can be written,
in the form
\begin{equation}
v'_{2}=v'_h + v'_p
\label{vseparation}
\end{equation}
where the indexes $h$ and $p$ stand for a solution of the homogenous
equation and a particular solution of the
inhomogeneous equation, respectively. As we have already remarked,
the thinness of the disk allows for the density and vertical velocity
fluctuations not to induce either radial or azimuthal motions at
these orders. This implies that homogeneous solutions of
(\ref{veq}), (i.e. with the equation's RHS set to zero,
because $u_1'=\Omega_2'=0$ say, see above) are perfectly
acceptable.

These VA solutions have the form (see UNRS)
\beq
v_h' = \hat v_h'(z,r)\exp(\sigma_{\rm v} T) + {\rm c.c.}~~~{\rm with}~~~
T\equiv{t \over r^{3/2}},
\label{homo_acoustics_ansatz}
\eeq
where the spatial eigenfunctions $\hat v_h'$ are composed of the associated Legendre functions.
This form of $v_h'$ already indicates that these solutions
are {\em inseparable} in $r$ and $t$.
The eigenvalues, which
appear in complex conjugate pairs, are functions
of $\alpha$ and in general they also depend, and quite sensitively,
on the index $n$ (see UNRS), but as mentioned before
we concentrate here on the case $n=3/2$ only.
The eigenfrequency of the fundamental mode,
for our chosen polytropic index is given by
\beq
\sigma_{\rm v} =
-\frac{4}{9}\alpha \pm  i \left| \left(\frac{16}{81}\alpha^2 -
\frac{8}{3} \right) \right| ^{1/2}.
\eeq
All such eigenfrequencies have a negative real part
and thus, as can be expected, the fundamental as well as all overtones,
show temporal decay.

\subsection{General acoustics (GA)}
\label{general} We turn now to the general solution to
(\ref{ueq}-\ref{veq}) by first focusing on
(\ref{ueq}).  This equation admits an infinite set of
eigenmode solutions in a similar way to the
homogeneous solutions discussed before.  The general
derivation and structure of these eigenmodes is detailed in
UNRS.  For the sake of clarity,
the discussion here focuses only upon the
dynamics associated with the fundamental mode of the operator
${\cal P}$. This solution is given by (see UNRS)
\beq
u_1' = \hat u'(z,r)\exp(\sigma_{\rm u} T + \vartheta) + {\rm c.c.},
\eeq
with $T\equiv{t/r^{3/2}}$, as before, and
in which the spatial eigenfunction $\hat u'(z,r)$ has the particularly simple structure
\beq
\hat u'(r,z) = A(r)\left[\frac{z^2}{h^2} - \frac{1}{6} \right],
\label{u_prime_solution}
\eeq
where $h$, the disk thickness is a known function of $r$, see (\ref{hr}).
$A(r)$ is an amplitude whose radial functional form is technically
arbitrary and would be set by the initial condition.
An arbitrary phase factor of $\vartheta$ has also been introduced. The
eigenvalues, denoted in this case by $\sigma_{u}$, again come in complex
conjugate pairs and the fundamental frequency is given by
\beq \sigma_{\rm u} =  - \sfrac{8}{5}\alpha \pm i.
\label{eigenvalue_inertioviscous_k0}
\eeq Without loss of
generality we may only consider  the `$+$' solution since the
phase factor $\vartheta$ can account for the `$-$' solution. The
temporal behavior of these modes is again one of decaying
oscillations, but with frequencies given by the local rotation
rate of the disk. Note that in the prescription for $h(r)$
as given in UNRS for values of $r$
sufficiently greater than the zero-torque radius, $r_*$,
(see KK) the
disk height with respect to $r$ is well approximated by,
\beq
h(r) \sim h_1 r,~~~ {\rm~and~thus}~~~ \frac{dh}{dr} \sim h_1,
\label{hr}
\eeq
where $h_1$ is determined by the mass flux rate and the value of $\alpha$.
Since $h_1$ simply adds a multiplicative factor to all the dynamical quantities, it plays no
role in the quality of the ensuing evolution and, as such, we set $h_1 = 1$
without any loss of general flavor.

With this solution for $u_1'$ we may find a particular solution to the vertical
velocity, $v_p'$ by solving (\ref{veq}), together with (\ref{ivp1}) directly.
The details of this, rather lengthy procedure are presented in UNRS, but we can highlight
the major steps here. We posit the following Ansatz
for $v_p'$ and $\rho_p'$,
\beqa
v_p' &=& \hat v_p (\zeta,r)e^{\sigma_{\rm u} T}
+ \hat V_{p}
(\zeta,r)\left(T e^{\sigma_{\rm u} T}\right) \ \ +  {\rm c.c.},
\label{vp_solution_form} \\
\rho_p' &=& \hat \rho_{p}(\zeta,r)e^{\sigma_{\rm u} T}
+ \hat R_{p}(\zeta,r)
\left(T e^{\sigma_{\rm u} T}\right) \ \ +  {\rm c.c.},
\label{rhop_solution_form}
\eeqa
where $\zeta \equiv z/h(r)$.

Recall that we solve only for the fundamental mode of the system (see
UNRS) and that there is obviously a dependence on $\alpha$. It is important to
notice the explicit appearance of a multiplicative $T$ term in the above expressions.
This temporal functional dependence is necessary in order to balance terms
arising on the RHS of (\ref{veq})- the `driving' term and will, of course,
cause TG (see below in the
next subsection).
In general, the form of these lowest order structure eigenfunctions is
\beq
\left(
\begin{array}{c}
\hat v_{p} \\
\hat V_{p}
\end{array}
\right)
=
\left(
\begin{array}{c}
a_1 \\
b_1
\end{array}
\right)\zeta
+
\left(
\begin{array}{c}
a_3 \\
b_3
\end{array}
\right)\zeta^3,
\eeq
namely, it is a polynomial in $\zeta$ of the third degree with only odd powers in $\zeta$.
The four coefficients are functions of $r$ and $\alpha$, i.e.
$a_i=a_i(r,\alpha)$ and $b_i=b_i(r, \alpha)$.

The solution to the density perturbation, $\hat \rho_p'$ can be quite straightforwardly obtained
using (\ref{u_prime_solution}), (\ref{vp_solution_form}) and  (\ref{rhop_solution_form}) , taking the functional
form of $\rho_0$ as detailed in the $\order1$  steady-state solution
(see KK, UNRS) and finally integrating
equation (\ref{ivp1}) with respect to time. This gives
\beqa
\left(
\begin{array}{c}
\hat \rho_{p} \\
\hat R_{p}
\end{array}
\right)
&=&
\bigl(1-\zeta^2\bigr)^{\sfrac{1}{2}} \times \nonumber\\
&& \left[
\left(
\begin{array}{c}
c_0 \\
d_0
\end{array}
\right)
+
\left(
\begin{array}{c}
c_2 \\
d_2
\end{array}
\right)\zeta^2
+
\left(
\begin{array}{c}
c_4 \\
d_4
\end{array}
\right)\zeta^4
\right],
\eeqa
where, again, the coefficients are functions of $r$ and $\alpha$, that is,
$c_i=c_i(r,\alpha)$ and $d_i=d_i(r, \alpha)$.

\par
The polynomials appearing in the square
brackets of the expressions for $ \hat \rho_{p}$ and $\hat R_{p}$
have only even powers of $\zeta$. The detailed forms of the coefficients
depend on the form of the generalized initial perturbation
radial structure $A(r)$ (as well as on $n$, in the general polytrope case).
We avoid presenting
them here because they are very long and cumbersome; and the details of the coefficients
have no effect on the
resulting transient dynamics, since they describe only a particular type of disturbance.
For the sake of simplicity
most analytic results which will be presented below assume $A = e^{i\pi/4}$
(that is, a particularly simple form of the initial conditions with
no dependence of the disturbance). We shall however show also one
instance of a more complicated  initial condition  consisting of
a localized Gaussian profile in $r$.

\subsection{Transient dynamics and growth}
\label{TG}
As is evident by inspection of (\ref{vp_solution_form}), $v'_p$ and
$\rho_p'$ exhibit a pronounced TG during a finite
time interval
(due to the factor $T$ multiplying the decaying exponential),
before eventually decaying.  To demonstrate this more clearly
we may consider some integrated energy quantities. The first
of these is an energy density, per unit {\em surface} of the disk
\beq
{\cal E}_a(r,t;\alpha) \equiv
h(r)\int_{-1}^{1}{\left(\sfrac{1}{2}\rho_0 v_p'^2 + \sfrac{1}{2}
\frac{c_{s0}^2\rho_p'^2}{\rho_0} \right)}d\zeta,
\label{energy_density_a_def}
\eeq
${\cal E}_a(r,t;\alpha)$ is to be interpreted as the
acoustic energy (per unit area of the disk) consisting of the
kinetic energy in the vertical velocity disturbances and the
compression energy due to the density disturbances.
The contribution of the (purely oscillatory)
velocity components, resulting from the homogeneous part of the
disturbance equations, are left out of (\ref{energy_density_a_def}).
\begin{figure}
\begin{center}
\leavevmode \epsfysize=5.cm
\epsfbox{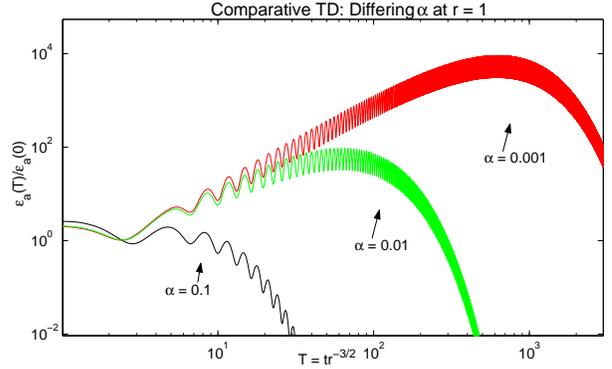}
\end{center}
\caption{{\small
The TG in the quantity ${\cal E}_a$ for the fundamental mode
at $r=1$, $n=3/2$ and where $A(r) = e^{i\pi/4}$, $dA/dr = 0$.  The four curves correspond to
$\alpha = 0.1, 0.01, 0.001$ and are presented
on a log-log plot.  All curves are ${\cal E}_a$, scaled to its
value at $T=0$.
The rise time, as well as the maximal value achieved, are proportional
to the inverse of $\alpha$.
${\cal E}_a$ also exhibits oscillations with period $\pi$ that
sit atop the general transient trend.
}}
\label{figure1}
\end{figure}

${\cal E}_a$  depends on  the $\alpha$ parameter
as well as the structure of the radial velocity perturbation $A(r)$
(cf. \ref{u_prime_solution}).
The choice of $A(r)$
as constant with respect to $r$ makes
the integral defined in (\ref{energy_density_a_def})
analytically tractable.
Nonetheless the expressions, which have been verified with the aid of Mathematica 5.0,
are very long and will not be displayed here;
only their essential features will be addressed.
With $A = e^{i\pi/4}$, ${\cal E}_a$ has the functional
form
\beq {\cal E}_a(r,t;\alpha)  = \frac{e^{-2 \alpha
T}}{r^{3/2}}\Upsilon(T,\cos{2T},\sin{2T};\alpha), \
\eeq
where
$\Upsilon$ is a well-defined analytical, albeit complicated,
function of its arguments. Thus,  ${\cal E}_a$  depends on time
only through the variable $T=t/r^{3/2}$, which is the time
measured in units of the local disk rotation period, at $r$,
divided by $2\pi$.

In Figure \ref{figure1}, the evolution of
${\cal E}_a$ as a function of the variable $T$ is shown
for different values of $\alpha$. $T$ is actually a similarity
variable and the value of $r$ explicitly appears only
in the pre-factor multiplying the function $\Upsilon$.
For the case displayed in the figure the radius
is fixed (set to $r=1$).
Different values of $r$ will merely change the
overall height of the response ($\propto r^{-(3/2)}$)
while the shape of the function is self-similar
and is always decaying at
long times $T$.
Inspection of the figure, uncovers a very important property of
the TG. As $\alpha$ decreases, the magnitude of the
growth becomes more prominent\footnote{For an inviscid disk, i.e.
with formally $\alpha=0$, Umurhan \& Shaviv (2006) obtained algebraic
growth with no decay, which is quite telling, but should be
considered carefully (see Sec. \ref{4sec} on the validity of
asymptotic expansions).}
 (for example, it is up to $\sim 3$ orders of
magnitude for $\alpha = 0.001$) and the maximum occurs at correspondingly
later times.
The time corresponding to the maximum amplitude $T_{\rm max}$ is
roughly $\sim 5/(8\alpha)$ in this\footnote{For the $k$-th
overtone, $T_{\rm max}$ will be smaller by a factor of $1/k^2$,
see UNRS)} case. The rise in energy is
modulated by oscillations which arise from the fact that there are
correlations between pairs of variables which contribute to the
overall growth in ${\cal E}_a(T)$. These variables
oscillate in a frequency defined by the imaginary part of the
eigenvalue (\ref{eigenvalue_inertioviscous_k0}) and
because ${\cal E}_a(T)$ is composed of
products of pairs of dynamical variables terms depending on $2T$ appear
- explaining why the observed period is half the orbital period at a given radius.

${\cal E}_a$ is a function of the radial position $r$, but
one may form the integral quantity $E_a(t;\alpha)$, i.e., the {\em total
disturbance acoustic energy} contained in a global portion of the disk
(a ring), by
integrating over the radial range in which these disturbances are
assumed to exist, that is, between the inner and outer bounds
$r_{\rm min}$ and $r_{\rm max}$ of the ring.
\beq
E_a(t;\alpha) \equiv \int_{r_{\rm min}}^{r_{\rm max}}{{\cal E}_a(r,t;\alpha) \
2\pi r \ dr}.
\label{energy_a_def}
\eeq
\begin{figure}
\begin{center}
\leavevmode \epsfysize=6.7cm
\epsfbox{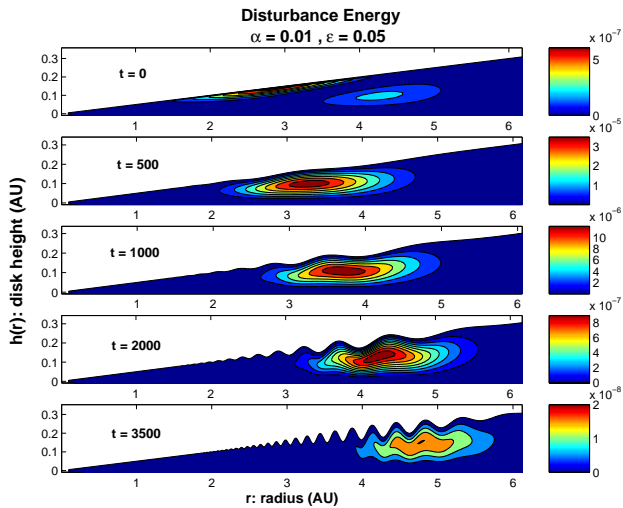}
\end{center}
\caption{{\small TG in the vertical velocity disturbance kinetic
energy density in a disk with $\alpha = 0.01, \epsilon =0.05$.
$A(r)$ is as in (\ref{gaussian}) with
$r_0 = 3$ and $\Delta = 2.5$. Time is given in terms of disk rotation times
at the radius $r=1$, so that $T=t$. The magnitude of the
disturbance energy density is indicated by the shadings
(whose meaning is indicated on the right of each panel) on a cut
through the disk's meridional plane. Significant growth in the
disturbance energy is observed (almost three orders of magnitude).
Crenellation patterns that appear in the fluctuating boundary of
the disk are ultimately wiped out by viscous decay.}}
\label{figure2}
\end{figure}
We have found (see UNRS) that $A(r)$ affects the
spatial details but not the global time behavior and
that the occurrence of the peak energy value is
delayed for larger values of $r_{\rm max}$, at given values of
$\alpha$ (and $r_{\rm min}$), although the value of the energy at the peak time
remains roughly the same for a given $\alpha$.
To get an idea about the expected qualitative {\em spatio}-temporal
behavior of such transiently growing solutions we have computed
the results for a case in which $A(r)$ is not constant (and, thus,
the possibility to evaluate the relevant integrals analytically
can not be expected, in general).  We consider here an initial condition
in which the radial velocity amplitude has a Gaussian form
around some fixed radius, $r_0$ say, and having a width $\Delta$,
such that it is well contained within the
region $r_{\rm min}\le r \le  r_{\rm max}$
\beq
A(r) = e^{i \pi/4}
\exp\left[-\frac{(r-r_0)^2}{\Delta_f}\right].
\label{gaussian}
\eeq
The fluid is assumed to be otherwise initially undisturbed.  As pointed out
above, with this form for $A$ we lose the ability to find an
analytic solution and must resort to numerical evaluation.
In Figure \ref{figure2} we show, in a contour plot in the
disk meridional plane, the kinetic energy (per unit volume)
contained in the vertical velocity disturbance $v_p$. The TG and
decay of the disturbance
is shown in the time sequence of figures displaying the spatial
structure of the disturbance. We also depict how the disk surface
moves in response to the imposed perturbation by solving the
equation of motion for the boundary at this order.
Time is given in units of rotation times of the disk, as measured
at the radius $r=1$.  Because the response of the disk surface is
an integral with respect to time,
it is not a surprise to see a large amplitude in the
surface position long after the kinetic energy has
started to die away.

\par
\section{Summary and Discussion}
\label{4sec}
We conclude with some remarks on possible improvements to
the asymptotic analysis of the sort done here and prospects for the future,
e.g., extensions to non-axisymmetric perturbations.
Asymptotic expansions, when viable, are often very robust and provide
a good approximation to the solution when truncation to only few leading
terms is done. Obviously, when a term in the series becomes {\em very} large
it may `break its order', that is, become larger than a previous term and
as such make the expansion invalid in this region. In our expansions
successive terms ratios are of $\order{\epsilon^2}$, and thus the procedure's
validity should not severely be limited even up to a growth factor of
$\sim 1000$ or so (in the velocity or density perturbations).
This matter is further discussed in UNRS, here we just remark
that the validity with respect to time for weakly viscous solutions are somewhat
influenced by the thinness of the disk:
smaller values of $\epsilon$ mean that the solutions are valid for longer times
after the initial disturbance.
More importantly, however, for a given value of $\epsilon$ one must
not be too zealous or overreaching by attempting to infer
the quantitative behavior of the disk
for arbitrarily small values of $\alpha$ - which, as one will recall, is formally
assumed here to be an $\order1$ quantity.

Despite these caveats,
the procedure when carried to higher order introduces corrections which are
technically non-linear.  Careful consideration must be undertaken in order to handle the
response at these higher orders.  This may entail treating the disturbance amplitudes for
the lower order solutions (like
$A(r)$ in $u_{1}'$) as {\em weakly non-linear} governed by a second `slow' time (e.g.
the amplitude is instead written as $A = A(r,\tau)$ where $\tau = \epsilon^2 t$) in a manner
analogous to the treatment of non-linear thick polytropes
(e.g, Balmforth \& Spiegel, 1996).

The approach used here may be generalized in a number of
directions. Allowing for non-axisymmetric
perturbations, including the disk inner and outer boundary in some
kind of boundary layer analysis and relaxing the polytropic
assumption seem to be the most obvious generalizations.  We have
found the presence of prominent TG in the simplest cases. It is difficult to
imagine that it will be suppressed in the more general conditions
although the effect of radiative energy losses on TG must be
carefully examined.

The question concerning the ultimate fate of the transiently grown
perturbations and their ability to induce a state of sustained complex dynamical
activity in the disk remains open. In this context it is worthwhile
to notice that since the TG decay times are of the order of
hundreds of rotation periods, it is conceivable that AD,
which are usually not isolated systems, may experience
recurrent external perturbations on such time scales and in this
way the dynamical activity may be sustained.
Extensive numerical calculations of
AD are however needed to decide if TG
may lead, through non-linear processes, to sustained turbulence,
or at least a dynamical state in which adequate angular momentum
transport can be sustained.
Such high-resolution global 3-D calculations are, however, still above
the ability of the present computer power and it may be thus advantageous to also exploit
sophisticated non-linear asymptotic methods to complement and
guide them.

\section{Afterword}
\label{5sec}
Accretion disks were introduced to explain astronomical X-ray sources
which were discovered close to four decades ago, when Jean-Pierre Lasota
was still a graduate student. At the time of the conference held at
the Trzebieszowice Castle, on the occasion of his 65-th birthday, it
seemed quite obvious that although we have gone a long way towards
understanding the physics of these fascinating and astronomically ubiquitous
objects, much still remains to be elucidated. In order to enable accretion
rates that are compatible with observations of such diverse objects as
forming stars, close binary systems or active galactic nuclei, an efficient
physical mechanism for angular momentum transport must be present in
these AD flows. It has been recognized at the outset that fluid
turbulence offers a natural mechanism of this sort, but hydrodynamical
(and also MHD) turbulence remains, unfortunately,
until this day to be one of the major bugbears of theoretical physics.

Shear flows are known to be particularly difficult in trying to understand
the physics of transition to turbulence in them. Even the simplest flows
of this kind, like the plane Couette and Poiseulle ones and the pipe flow,
are not well understood in this respect (see, e.g., Drazin 2002). How, then,
can we reasonably expect to adequately understand this problem in AD,
with their literally astronomical $\rm Re$, strong shear and rotation,
compressibility, magnetic fields permeating an ionized medium and other
physical complications?
The lesson learned from stellar turbulent convection is that the crudest
(e.g, mixing-length) phenomenological approaches may be very effective
in constructing observationally viable models of stars, but detailed
understanding is extremely difficult.

As far as AD are concerned, making significant progress beyond
the famous $\alpha$ model requires, at least in my view, a combined synergetic
effort of different groups of researchers, approaching the problem
from many different angles and employing essentially all relevant methods that
theoretical physics can offer. Phenomenology, analytical and semi-analytical
investigation of simplified systems, laboratory experiments and numerical
simulations of various kinds can all be useful, and especially if they
fertilize each other. Paradigms have always played an important r\^ole
in scientific progress, often consecutively replacing one another.
Dogmas, however, have always been stumbling blocks, especially in the
education and mentoring of young researches who are naturally able
to offer fresh and independent ideas.

If we are able to avoid such dogmas and truly cooperate in the way hinted on
above, there are good chances that the physics of accretion disks will be much
better understood when we celebrate Jean-Pierre's proverbial
120-th birthday, and hopefully even much sooner.\\

\leftline{\sl Acknowledgements}
\medskip
\small

\noindent
I thank Bruno Coppi, Wlodek Klu\'zniak, Mario Livio, P.-Y. Longaretti,
Miki Mond, Giora Shaviv, Frank Shu, Ed Spiegel,
Phil Yecko and Jean-Paul Zahn for discussing some problems related to the research described
here and for their encouragement. I am particularly indebted to my young collaborators
and especially to Orkan Umurhan, for committing his talents and investing a lot of hard work
in this endeavor. Finally, I would like to thank Marek Abramowicz for his hard work in organizing
this successful conference.






\begin{thebibliography}{}

\small

\bibitem{afshordi}
      Afshordi, N., Mukhopadhyay, B. \& Narayan, R. 2005, ApJ, 629, 383

\bibitem{bal03}
       Balbus, S. A. 2003, ARA\&A, 41, 555

\bibitem{bh91}
       Balbus, S. A. \& Hawley, J. F. 1991, ApJ, 376, 214

\bibitem{bh98}
       Balbus, S. A., \& Hawley, J. F. 1998, Rev. Mod. Phys., 70, 1

\bibitem{bhs96}
       Balbus, S. A., Hawley, J. F., \& Stone, J. M. 1991, ApJ, 476, 76

\bibitem{bs96}
          Balmforth, N.J. \& Spiegel, E.A., 1996, Physica D, 97, 1

\bibitem{barranco}
       Barranco, J. \& Marcus, P.S. 2005, ApJ, 623, 1157

\bibitem{bodo}
      Bodo, G., Tevzadze, A., Chagelishvili, G., Mignone, A., Rossi, P.
       \& Ferrari, A. 2007, A\&A, 475, 51

\bibitem{spiegel}
        Bracco, A., Chavanis, P. H., Provenzale, A. \& Spiegel, E. A. 1999,
        Phys. Fluids, 11, 2280

\bibitem{chagelishvili03}
     Chagelishvili, G. D., Zahn, J.-P., Tevzadze, A. G. \& Lominadze, J. G. 2003,
      A\&A  402, 401-407

\bibitem{coppi03}
     Coppi, B. \& Keyes, E. A. 2003, ApJ, 595, 1000

\bibitem{dubrulle}
      Dubrulle B., Marie L. , Normand Ch., Richard D. ,  Hersant F.\&  Zahn  J.-P.
      2005, A\& A, 429, 1


\bibitem{chandra60}
       Chandrasekhar, S. 1960, Proc. Natl. Acad. Sci., 46, 53

\bibitem{criminale}
    Criminale, W.O., Jackson, T.L. \& Joslin, R.D. 2003, {\em Theory and Computation
    in Hydrodynamic Stability}, Cambridge Univ. Press, Cambridge

\bibitem{drazin}
       Drazin, P. G. 2002, {\em Introduction to Hydrodynamic Stability}, Cambridge Univ.
       Press, Cambridge

\bibitem{fkr}
      Frank J., King A.R. \& Raine D.J. 2002, {\em Accretion Power in Astrophysics}, Cambridge
      Univ. Press, Cambridge


\bibitem{pap1}
        Fromang, S., \& Papaloizou, J. 2007, A\&A, 476, 1113

\bibitem{pap2}
        Fromang, S., Papaloizou, J., Lesur, G., \& Heinemann, T.
        2007, A\&A, 476, 1123


\bibitem{gl}
       Godon, P. \& Livio, M. 1999, ApJ, 523, 350

\bibitem{glb65}
       Goldreich, P., \& Lynden-Bell, D. 1965, MNRAS, 130, 125


\bibitem{grossman}
      Grossman, S. 2000 Rev. Mod. Physics, 72, 603


\bibitem{jg}
      Ji, H., Burin, M., Schartman, E. \& Goodman, J. 2006, Nature, 444, 343


 \bibitem{ioannou01}
     Ioannou, P.J. \& Kakouris, A. 2001, ApJ,  550, 931

 \bibitem{kpl07}
      King, A. R., Pringle, J. E., \& Livio, M. 2007, MNRS, 376, 1740

\bibitem{klahr03}
     Klahr, H.H. \& Bodenheimer, P. 2003, ApJ, 582, 869


\bibitem{kluzniak}
      Klu\'zniak W. \& Kita, D. 2000, Three-dimensional structure of an
         alpha accretion disk, {\tt arXiv:astro-ph/0006266v1} (KK)


 \bibitem{ll05}
      Lesur, G., \& Longaretti, P.-Y. 2005, A\&A, 444, 25

 \bibitem{ll07}
      Lesur, G., \& Longaretti, P.-Y. 2007, MNRAS, 378, 1471

 \bibitem{colgate}
     Li, H., Colgate, S. A., Wendroff, B. \& Liska, R. 2001, ApJ, 551, 874

 \bibitem{lp}
    Lin, D.N.C. \& Papaloizou, J.C.B. 1996, ARA\&A, 34, 703

 \bibitem{lithwick}
    Lithwick, Y. 2007, Formation, Survival, and Destruction of Vortices in Accretion Disks,
                       {\tt arXiv:0710.3868v1 [astro-ph]}

\bibitem{Mukhopadhyay}
      Mukhopadhyay, B., Afshordi, N. \& Narayan, R. 2005, ApJ, 629, 383


 \bibitem{pcp07}
      Pessah, M. E., Chan, C-k. \& Psaltis, D. 2007, ApJ , 668, L51

\bibitem{petersen}
      Petersen, M. P., Stewart, G. R. \& Julien, K. 2007, ApJ , 658, 1252

 \bibitem{pb68}
      Prendergast, K. H. \& Burbidge, G. R., 1968,
      ApJ, 151, L83

 \bibitem{pr72}
      Pringle, J. E. \& Rees, M. E. 1972, A\&A, 21, 1

\bibitem{pringl}
      Pringle, J.E., 1981, ARA\&A, 19, 137

\bibitem{regev83}
      Regev, O.  1983,  A\&A,  126, 146

\bibitem{bertout95}
      Regev, O. \& Bertout, C. 1995, MNRAS,  272, 71

\bibitem{gitelman}
      Regev, O. \& Gitelman, L. 2002, A\&A,  396, 623

 \bibitem{roc07}
      Rincon, F., Ogilvie, G. I. \& Cossu, C.  2007, A\&A,   463, 817


\bibitem{schmid01}
      Schmid, P.J. \& Henningson, D.S. 2001, {\it Stability and Transition in Shear Flows},
      Springer, New York

\bibitem{ss73}
      Shakura, N. I., \& Sunyaev, R. A. 1973, A\&A, 24, 337

\bibitem{shu}
      Shu, F. H., Galli, D., Lizano, S.,; Glassgold, A. E.; Diamond, P. H. 2007,
      ApJ, 665, 535


\bibitem{tassoul}
       Tassoul, J.-L.  1978, {\em Theory of Rotating Stars}, Oxford University Press, Oxford.

\bibitem{tevzadze03}
   Tevzadze, A. G., Chagelishvili, G. D., Zahn, J.-P., Chanishvili, R. G. \& Lominadze, J. G. 2003,
   A\&A,  407, 779

\bibitem{toomre}
     Toomre, A. 1964, ApJ, 139, 1217

\bibitem{umurhan06}
       Umurham, O. M., 2006, MNRAS, 365, 85

\bibitem{ur04}
       Umurhan, O.M. \& Regev, O., 2004, A\&A, 427, 855

\bibitem{us05}
        Umurhan, O.M. \& Shaviv, G. 2005, A\&A, 432, L31

\bibitem{umr07}
     Umurhan, O. M., Menou, K. \& Regev, O. 2007, Phys. Rev. Lett.,
     98, 034501

\bibitem{Nemi}
    Umurhan, O. M, Nemirovsky, A., Regev, O. \& Shaviv, G. 2006,
    A\&A, 446, 1 (UNRS)


\bibitem{urm07}
     Umurhan, O. M., Regev, O. \& Menou, K. 2007, Phys. Rev. E, 76, 036310


\bibitem{vel59}
      Velikhov, E. P. 1959, J. Exp. Theor. Phys., USSR, 36, 1398

\bibitem{yecko04}
      Yecko, P.A. 2004, A\&A,  425, 385

\end{thebibliography}
\end{document}